\documentclass[sigconf]{acmart}
\usepackage[ruled,vlined]{algorithm2e}
\usepackage{mathtools}
\usepackage{float}
\usepackage{placeins}
\usepackage{diagbox}

\DeclarePairedDelimiter\floor{\lfloor}{\rfloor}
\AtBeginDocument{%
  \providecommand\BibTeX{{%
    \normalfont B\kern-0.5em{\scshape i\kern-0.25em b}\kern-0.8em\TeX}}}

\setcopyright{acmcopyright}
\copyrightyear{2020}
\acmYear{2020}
\acmDOI{10.1145/1122445.1122456}

\acmConference[WSDM '21]{WSDM '21: ACM Symposium on Neural
  Gaze Detection}{March 08--12, 2021}{Jerusalem, Israel}
\acmPrice{15.00}
\acmISBN{978-1-4503-XXXX-X/18/06}

\begin{document}

%%
%% The "title" command has an optional parameter,
%% allowing the author to define a "short title" to be used in page headers.
\title{CONQ: \underline{CON}tinuous \underline{Q}uantile Treatment Effects for Large-Scale Online Controlled Experiments}

%%
%% The "author" command and its associated commands are used to define
%% the authors and their affiliations.
%% Of note is the shared affiliation of the first two authors, and the
%% "authornote" and "authornotemark" commands
%% used to denote shared contribution to the research.
\author{Weinan Wang}
\email{wwang@snapchat.com}
\affiliation{%
  \institution{Snap Inc.}
  \streetaddress{2850 Ocean Park Blvd.}
  \city{Santa Monica}
  \state{California}
  \country{USA}
}

\author{Xi Zhang}
\email{xzhang2@snapchat.com}
\affiliation{%
  \institution{Snap Inc.}
  \streetaddress{2850 Ocean Park Blvd.}
  \city{Santa Monica}
  \state{California}
  \country{USA}}

\renewcommand{\shortauthors}{Wang and Zhang, et al.}

\begin{abstract}
In many industry settings, online controlled experimentation (A/B test) has been broadly adopted as the gold standard to measure product or feature impacts. Most research has primarily focused on user engagement type metrics, specifically measuring treatment effects at mean (average treatment effects, ATE), and only a few have been focusing on performance metrics (e.g. latency), where treatment effects are measured at quantiles. Measuring quantile treatment effects (QTE) is challenging due to the myriad difficulties such as dependency introduced by clustered samples, scalability issues, density bandwidth choices, etc. In addition, previous literature has mainly focused on QTE at some pre-defined locations, such as P50 or P90, which doesn't always convey the full picture. In this paper, we propose a novel scalable non-parametric solution, which can provide a continuous range of QTE with point-wise confidence intervals while circumventing the density estimation altogether. Numerical results show high consistency with traditional methods utilizing asymptotic normality. An end-to-end pipeline has been implemented at Snap Inc., providing daily insights on key performance metrics at a distributional level. 
\end{abstract}

\maketitle

\section{Introduction}
Large-scale online experiments \citep{Box05,Ger12} are routinely conducted at various technology companies (Google, Netflix, Microsoft, LinkedIn, etc.) \citep{Xie16,Koh09,Tang10,Bak14,Xu15} to help gauge quantifiable impacts of potential feature and product launches. At Snap Inc., hundreds of experiments are conducted on a daily basis, aiming to improve user engagement or app-performance. To accurately measure these changes, hundreds and thousands of metrics are further pumped through our in house A/B platform where routine statistical tests are conducted to judge if the treatment effects are indeed statistically significant. These metrics can be roughly divided into two types: engagement vs. performance. For engagement type metrics, typically two sample $t$-tests are utilized to determine if the average treatment effects (ATE) are truly nonzero \citep{Imb15}. However for performance metrics (e.g., camera transcoding latency, page load latency, etc.), evaluating treatment effects at mean is typically ill-advised \citep{Liu19}, and the industry standard is to measure QTE (quantile treatment effects), most notably at P50 and P90, whereas P50 is aimed at a summary statistic for overall performance and P90 for tail area performance. 

\begin{figure}[!h]
    \centering
    \includegraphics[scale=0.31]{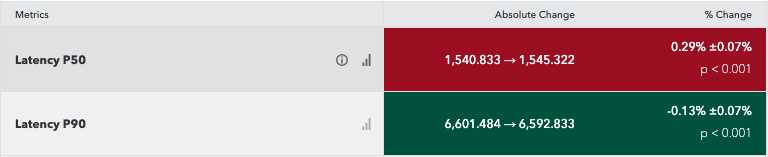}
    \caption{A sample experiment where a performance metric has statistically significant QTE at both P50 and P90, however the direction differs.}
    \label{fig:quantile_example}
\end{figure}

However, it is not straightforward to get such QTE with valid $p$-values and confidence intervals as events can no longer be considered independent (each randomization unit, typically users, can contribute multiple events, which in itself is another random variable measuring engagement), therefore delta-methods and user-level correlation calculations are required \citep{Liu19}. In addition, while the method proposed in \citep{Liu19} is statistically valid, it lacks a principled and data-driven way on choosing the bandwidth for its kernel density estimation, which affects the result greatly since the density's square appears as the denominator of the variance term. Furthermore, QTE at the median and the $90$-th quantile are sometimes not enough to give experimenters the whole picture, especially when the significance or directions do not agree (example in Figure \ref{fig:quantile_example}), or when heterogeneous treatment effect (HTE) is present for different devices with high and low overall performances. This motivates us to come up with a robust, statistically valid, and scalable method which can provide continuous QTE with point-wise $p$-values and confidence intervals.

In this paper, we propose a novel method named CONQ (continuous quantile treatment effect), which utilizes Bag of Little Bootstrap (BLB) \citep{Klei14} and the Bahadur Representation \citep{Wu05} to construct point-wise asymptotically valid confidence intervals and $p$-values, which are consistent with the normal approximation method in \citep{Liu19}. To further improve Monte Carlo efficiency, balanced bootstrap is implemented for variance reduction. We also show the added benefit of log-transformation on the original metric for calculating percentage change QTE due to quantile's invariance to monotone transformations. Extensive evaluation on Snapchat data shows high consistency of our method with method in \citep{Liu19} at P50 and P90, and stable performance across a range of quantile locations bench-marked by known A/A tests. A daily pipeline has been implemented using Spark \citep{Zaha10} and Airflow, generating continuous QTE on core performance metrics for all A/B tests at Snap Inc.

The key contributions of our paper include:
\begin{itemize}
    \item 
    A scalable and theoretically sound method that can provide QTE with percentage change $p$-values and confidence intervals at arbitrary range of quantile locations simultaneously.
    \item
    Circumvents the issue of density estimation and bandwidth choice altogether, improve accuracy of delta-method by log-transformation for percentage changes.
    \item
    Validation of the method on real experiments at Snap Inc., which shows consistency with existing P50 and P90 results, and stable performance across various quantile locations.
\end{itemize}
In Section 2, we describe the background and literature on QTE in the A/B setting and statistical inference for quantiles. In Section 3, we detail the motivation and theoretical justification of the CONQ procedure and its implementation. In Section 4, we evaluate CONQ using a large set of A/B experiments at Snap Inc., showing its comparable performance with delta-method based approach in \cite{Liu19, Deng18}, and shows its stable performance using a set of known A/A tests as validation corpus. In Section 5, we discuss the possible extension and summarize the paper.

\section{Background and Related Work}
\subsection{Quantile Treatment Effects}
Although in many tech firms, the vast majority of metrics are measuring user engagement activities, another important category of metrics aim to gauge app or web performance, i.e. performance metrics. For such performance metrics, not only do we care about the average users' experience, we are also interested in the cohort of users that are suffering from the worst app or web performances. Bearing these concerns in mind, it's important to evaluate these metrics at at least multiple quantile locations in A/B tests, especially in the tail area. At Snap Inc., P50 and P90 are the typical objectives various teams optimize for in A/B experiments. \citep{Liu19} proposed a statistically valid and scalable method that can calculate QTE with $p$-values and confidence intervals by appealing to quantile estimator's
asymptotic normality \cite{Cramer99}, however the varying bandwidths choice mentioned is rather ad-hoc and lacks theoretically guarantee. \citep{Deng18} introduced an approach that appeals to delta-method and the outer CI, circumventing the density estimation which still involves user-level correlation calculation for the variance term. Furthermore, these approaches still focus on treatment effects at a number of pre-determined quantile locations (.e.g., P50 or P90), where generalizations to a continuous range of quantile locations cannot be easily obtained without point-wise recalculation.

QTE at various quantile locations has been receiving increased attention these days in the industry, as it provides a fuller picture of distributional changes between control and treatment, and better captures heterogeneity \citep{Lux18}. At Uber, the approach they adopted is quantile regression, citing the advantage of relying on existing abundant literature \citep{Koe01,Hunter00}. However, the methodology proposed still requires an optimization algorithm, which is not suitable for the myriad of performance metrics and experiments at Snap Inc. at scale.  
\subsection{Statistical Inference for Quantiles}
\label{ref:quantile}
In the Statistics literature, the relationship between quantiles and CDF (cumulative distribution function) has been investigated extensively. Most notably, for $X_1,X_2,\cdots,X_n$ that are i.i.d. random variables from an unknown non-parametric $F$ with density $f$, for $0<p<1$, denote by $\xi_p=\inf\{x:F(x)\geq p\}$ the $p$-th quantile of $F$, $\xi_{n,p}$ the $p$-th sample quantile of $F_n$,  \citep{Baha66} coined the famed Bahadur representation:
$$
\xi_{n,p}=\xi_p+\frac{p-F_n(\xi_p)}{f(\xi_p)}+O_{a.s.}[n^{-3/4}(\log n)^{1/2}(\log\log{n})^{1/4}].
$$
To put it simply, it stated that an asymptotic relationship between quantile and CDF can be established through its density function $f$. Refinements of Bahadur's result in the i.i.d. setting were further proposed by Kiefer \citep{Kie67,Kie70,Kie70old}. \cite{Wu05} further extended such representation for a wide range of dependent cases. \cite{Wood52} proposed the Woodruff confidence interval, which 
essentially established a one to one relationship between point-wise confidence intervals of the CDF and quantiles.

Another line of research focused on a $\mathcal{L}_\infty$-norm deviation of the empirical CDF from the truth (uniform bound), instead of at a specific quantile location (point-wise). The celebrated Dvoretsky-Kiefer-Wolfowitz-Massart (DKWM) inequality \citep{Dvo56,Massart90} provides a tight bound on the CDF for i.i.d. samples:
$$
\mathbb{P}\left(\sup_{x\in\mathbb{R}}\left(F_n(x)-F(x)\right)>\epsilon\right)\leq \exp^{-2n\epsilon^2}, \;\forall \epsilon\geq \sqrt{\frac{1}{2n}\ln{2}}
$$
 , which by its expression, is indifferent to quantile locations. \citep{Ram19} further proposed two methods for point-wise quantile confidence intervals and two methods for $\mathcal{L}_\infty$-norm type confidence bounds, which are proven to be adaptive in a sequential setting, resonating with the always-valid $p$-values line of research \citep{Jo15,Joha17}. However, coming from a QTE perspective, we are still most interested in point-wise statistical inference, and \citep{Ram19}'s method requires either parameter tuning or numeric root finding, and generalization to the dependent case is non-trivial.

In this paper, we focus on the end-horizon time setting which is the common-case at Snap Inc., instead of a sequential test setting as considered by \cite{Ram19}. We propose an easy to interpret and implement data-driven method for a range of quantile level treatment effects for percentage changes.
\section{Continuous point-wise Statistical Inference on QTE}
\subsection{Sample Quantiles and Bahadur Representation}
At Snap Inc., hundreds of metrics are evaluated in A/B experiments simultaneously, aiming to not only gauge user telemetry data, but also on app-performance. As a social network company with a camera focus, myriad facets of in-app performances are the direct goal of optimization for many teams. There's almost always a dual performance metric for an engagement metric. For example, as we measure the number of app opens, we also measure app open latency. 

\begin{figure*}[!h]
    \centering
    \includegraphics[scale=0.7]{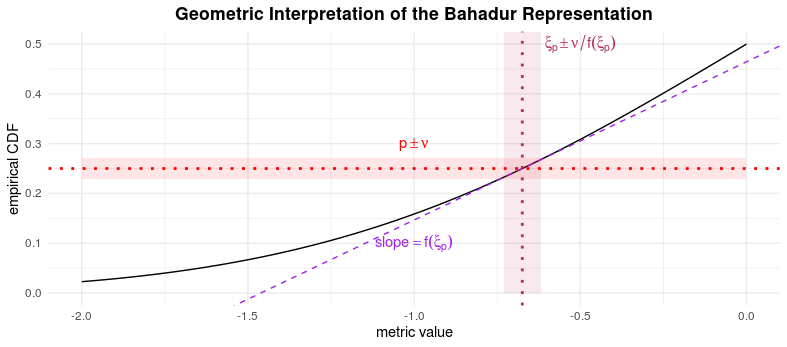}
    \caption{Geometric interpretation of the Bahadur representation between the empirical CDF and metric quantiles.}
    \label{fig:bahadur}
\end{figure*}

This brings about another layer of complication in A/B tests. As a typical practice, randomization units are users; for engagement type metrics, each user contributes exactly one point, and all data-points can be considered i.i.d. observations, hence statistical inference can be by and large accomplished by appealing to the Central Limit Theorem. However, the dual performance metrics are not as straight-forward, as each user contributes multiple data-points (exactly the engagement count number of data-points), therefore we have a clustered set of observations which are no longer independent.  
Using the same notation in \ref{ref:quantile}, for i.i.d. samples $X_1,\cdots,X_n$, the sample $p$-th quantile $\xi_{n,p}$ is asymptotically normal:
$$
\sqrt{n}\left(\xi_{n,p}-\xi_p\right)\rightarrow \mathcal{N}\left(0,\frac{p(1-p)}{f\left(\xi_p\right)^2}\right).
$$

With performance metrics, typically we need delta method to approximate the numerator in the variance term \cite{Deng18,Liu19}. However, this is only for a fixed quantile location $p$, and it's hard to generalize to other locations without recalculation. Furthermore, kernel density estimation is typically required for $\hat{f}(\xi_p)$, in which its bandwidth is hard to choose. Although some ad-hoc rules could work well in practice \cite{Liu19}, small deviations in the bandwidth affect the result greatly.

To demonstrate such issue, here is an example on a typical performance metric measuring the latency on starting the Discover section inside the Snapchat app. The study improved this latency at P50 from $911.991$ to $908.173$, with a $p$-value of $0.044$ using the method mentioned in \cite{Liu19} with the recommended bandwidth choice. However, if we were to vary the bandwidth using the below choices (with normal reference rule \citep{Sil86} chosen as $h_n=1.06\min\{s,\text{IQR}/1.34\}/n^{1/5}$ (IQR iinter-quantile range, $s$ is sample standard deviation and $n$ is sample size), we get drastically different $p$-values demonstrated in Table \ref{tab:density}.

\begin{table}[!h]
\begin{tabular}{|cc|}
\hline
\textbf{kernel density estimator's bandwidth}    & $\Delta\%$ $p$-\textbf{value} \\ \hline
0.002                 & 1.08e-34           \\ \hline
0.01                  & 0.0125             \\ \hline
0.02                  & 0.049              \\ \hline
normal reference rule & 0.352              \\ \hline
\end{tabular}
\caption{\label{tab:density}How bandwidth choice affects kernel density estimation, resulting in different percentage change $p$-values at P50.}
\end{table}
Using $0.05$ as a threshold, we can see that $p$-values vary from being very significant to not significant at all as the bandwidth varies. However, we lack a robust and data-driven approach on choosing such bandwidth which can guarantee good estimation quality for inference across all quantiles.

As noted in sub-Section \ref{ref:quantile}, the famed Bahadur representation establishes the asymptotic relationship between sample quantiles and the empirical CDF, especially in various dependent cases \cite{Wu05}.

Such Bahadur representation has a nice geometric interpretation as illustrated by \ref{fig:bahadur}. At the neighborhood of point $(\xi_p,p)$, to convert from the confidence interval of empirical CDF to quantiles, we can simply divide by the tangent of the curve $F(\xi_p)$, which is $F'(\xi_p)=f(\xi_p)$. Therefore, for i.i.d. samples, we have
$$
p-F_n\left(\xi_p\right)\rightarrow \mathcal{N}\left(0,\frac{p(1-p)}{n}\right)
$$
which no longer involves the density estimation. For clustered sample's case, simply replacing the variance term $p(1-p)/n$ with $\nu$ would suffice, which $\nu$ can be empirically estimated using the Bag of Little Bootstrap (BLB) technique in the next section. Take one step further, we have the result that:
\begin{align*}
\text{Let } \hat{\xi}_p^L&=\inf\left\{t:F_n(t)\geq p-z_{\alpha/2}\nu^{1/2}[F_n(\xi_p)]\right\},\\
\hat{\xi}_p^U&=\inf\left\{t:F_n(t)\geq p+z_{\alpha/2}\nu^{1/2}[F_n(\xi_p)]\right\},\;\text{then }\\
\mathbb{P}\left(\hat{\xi}_p^L\leq \xi_p\leq \hat{\xi}_p^U\right)&\approx 1-\alpha.
\end{align*}
Woodruff \citep{Wood52} first proposed this interval empirically, which ingeniously circumvents the density estimation altogether, while still provides valid statistical inference results.

Below we discuss how we extend such CI to a continuous setting and clustered sample's case, suitable for the large-scale A/B testing need at Snap Inc.

\subsection{Balanced Bag-of-Little-Bootstrap for CDF}
In the previous section, we discuss the relationship between empirical CDF and quantiles through the Bahadur representation. In this section, we discuss how the variance term $\nu$ can be efficiently calculated for at a continuous range of quantile locations using one set of bootstrap results, and how Woodruff type confidence intervals can be applied.

As mentioned in \cite{Deng18}, 'ntiles' operation (query quantile from set of observations) is expensive and should be avoided if possible. Furthermore, bearing in mind the one to one relationship between CI for quantiles and CI for CDF, in order to simultaneously get quantile level inference for all possible locations, we can first tackle the variance of CDF, and then appeal to Woodruff CI for transforming back to quantiles. In this paper, we use the Bag of Little Bootstrap in \citep{Klei14} with the balanced method \citep{Glea88} for resampling to further reduce Monte Carlo variance, achieving higher bootstrap efficiency.

Suppose there are $N$ users, for user $i=1,\cdots,N$, there are total of $M_i$ events, denoted as $X_{i,1},\cdots,X_{i,M_i}$, while $\sum_{i=1}^NM_i=n$. We first take the logarithm of these metrics, then preserve a pre-defined digits of precision (we chose 2): $\text{round}(\log{X_{i,M_i}},2),\;i=1,\cdots,N$. Note this step is to reduce the total number of unique log-scaled values for efficiency. Furthermore, due to quantile's invariance to monotone transformation \citep{Berg13}, we have
$$
\log\xi_{f(x)}(p)=\xi_{f(\log x)}(p),\;\forall 0<p<1.
$$ This ensures we can get back the QTE on the original metric $X$ by simply taking the exponent. Furthermore, by delta method, we have
\begin{align*}
&\text{Var}(\%\text{diff between control and treatment at p-th quantile})\\
=&\text{Var}\left(\frac{\xi^T_{f(x)}(p)-\xi^C_{f(x)}(p)}{\xi^C_{f(x)}(p)}\right)=\text{Var}\left(\frac{\xi^T_{f(x)}(p)}{\xi^C_{f(x)}(p)}\right)\\
=&\text{Var}\left(\exp\left(\log\xi^T_{f(x)}(p)-\log\xi^C_{f(x)}(p)\right)\right)\\
=&\text{Var}\left(\exp\left(\xi^T_{f(\log{x})}(p)-\xi^C_{f(\log{x})}(p)\right)\right)\\
\approx & \text{Var}\left(\xi^T_{f(\log{x})}(p)-\xi^C_{f(\log{x})}(p)\right)\left(\frac{\xi^T_{f(x)}(p)}{\xi^C_{f(x)}(p)}\right)^2\\
=&\left(\text{Var}\left(\xi^T_{f(\log{x})}(p)\right)+\text{Var}\left(\xi^C_{f(\log{x})}(p)\right)\right)\left(\frac{\xi^T_{f(x)}(p)}{\xi^C_{f(x)}(p)}\right)^2,
\end{align*}
where the first term in the last expression can be bootstrapped on log-transformed data separately for control and treatment. This way of applying the delta-method is empirically more accurate than on the original scale metric, especially when percentage change is large, or when $\xi^C_{f(x)}(p)$ is small.

Then we split $N$ users together with their corresponding events into $s$ subsets ($s=100$ for Snap) for control and treatment separately, all events for any given user shall be preserved in the same subset. For each subset, generate a data sketch with unique rounded log-scaled metric values (denoted by $\tilde{X}_j,\;j=1,\cdots,K$, $K$ here is the total number of unique values in the original dataset) with its corresponding count $c_{ij},\;i=1,\cdots,s,\;j=1,\cdots,K$:
\begin{table}[!h]
\caption{\label{tab:datasplit}Disjoint subsets splitting the original data, events belong to the same user are in the same bucket.}
\begin{tabular}{|c|l|l|l|l|}
\hline
 & \multicolumn{1}{c|}{\textbf{1}} & \multicolumn{1}{c|}{\textbf{2}} & \multicolumn{1}{c|}{\textbf{$\cdots$}} & \multicolumn{1}{c|}{\textbf{s}} \\ \hline
$\tilde{X}_{1}$                                & $c_{1,1}$                         & $c_{2,1}$                         & $\cdots$                               & $c_{s,1}$                         \\ \hline
$\tilde{X}_{2}$                                & $c_{1,2}$                         & $c_{2,2}$                         & $\cdots$                               & $c_{s,2}$                         \\ \hline
$\cdots$                                     & $\cdots$                        & $\cdots$                        & $\cdots$                               & $\cdots$                        \\ \hline
$\tilde{X}_{K}$                                & $c_{1,K}$                         & $c_{2,K}$                         & $\cdots$                               & $c_{s,K}$                         \\ \hline
\end{tabular}
\end{table}

Let $B$ be the number of bootstraps, to achieve balanced bootstrapping, i.e. where all samples appear exactly $B$ times, randomly permute the long vector
$
\pmb{V}=\left\{\underbrace{1,\cdots,1}_\text{B repetition},\underbrace{2,\cdots,2}_\text{B repetition},\cdots\cdots,\underbrace{s,\cdots,s}_\text{B repetition}\right\}
$
and then split $\pmb{V}$ into $B$ vectors of length $s$, and treat each vector $\pmb{v}_i,\;i=1,\cdots,B$ as one bootstrap sample on the buckets. For each bootstrap sample vector $\pmb{v}_i$, we can get a counter of how many times $1,\cdots, s$ each appears, and weight corresponding $c_{i,j}$ by how many times $i$-th bucket appears in $\pmb{v}_i$. For each bootstrap sample, we can get the empirical cdf at all unique log-scaled values $\tilde{X}_j,\;j=1,\cdots,K$. Using all bootstrap samples, we can get an estimate of the standard deviation of empirical cdf at all $\tilde{X}_j$ (denoted as $\mathbb{F}_{n,i}(\tilde{X}_j),\;i=1,\cdots,B$) as well, which would approximate $v$ at $\mathbb{F}_n(\tilde{X}_j)$: 
\begin{align*}
\text{Var}\left(\mathbb{F}_n(\tilde{X}_j)
\right)&\approx B^{-1}\sum_{i=1}^B \left(\mathbb{F}_{n,i}(\tilde{X}_j)-B^{-1}\sum_{k=1}^B\mathbb{F}_{n,k}(\tilde{X}_k)\right)^2\\
&:=\hat{\nu}^B_{n}[\mathbb{F}_n(\tilde{X}_j)].
\end{align*}

Bootstrapping on the empirical cdf avoids the 'ntiles' operations, and enables us to use one set of bootstrap to get all potentially interested QTE by interpolating the estimated variance via a general continuity assumption. Below is a typical example of the interpolated bootstrapped standard error across all quantiles:
\begin{figure}[!h]
    \centering
    \includegraphics[scale=0.38]{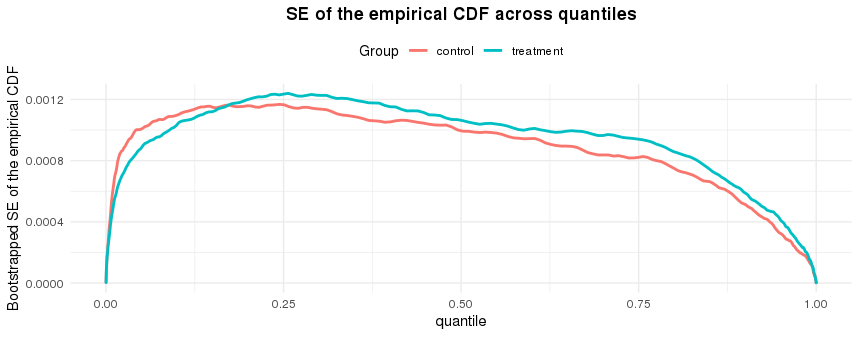}
    \caption{A typical example of bootstrapped SE for the empirical CDF across all quantile locations for control and treatment.}
    \label{fig:est_se}
\end{figure}

Since the original Woodruff method does not always produce a symmetric CI, empirically, we propose the following conservative estimate for each unique log-scaled metric value $\tilde{X}_j$ for control and treatment separately:
\begin{align}
\text{Let }\hat{\xi}_{\mathbb{F}_n(\tilde{X}_j)}^L&=\inf\left\{t:\mathbb{F}_n(t)\geq \mathbb{F}_n(\tilde{X}_j)-\hat{\nu}^B_n[\mathbb{F}_n(\tilde{X}_j)]\right\},\\
\hat{\xi}_{\mathbb{F}_n(\tilde{X}_j)} ^U&=\inf\left\{t:\mathbb{F}_n(t)\geq \mathbb{F}_n(\tilde{X}_j)+\hat{\nu}^B_n[\mathbb{F}_n(\tilde{X}_j)]\right\},\\ \text{then let }
\widehat{\text{SE}}(\tilde{X}_j)&=\max\left\{\tilde{X}_j-\hat{\xi}_{\mathbb{F}_n(\tilde{X}_j)}^L,\hat{\xi}_{\mathbb{F}_n(\tilde{X}_j)}^U-\tilde{X}_j\right\}.
\end{align}

\begin{algorithm*}[!thp]
\caption{\textbf{CONQ}: Continuous QTE on performance type metrics for large-scale online controlled experiments.}\label{CONQ}
\textbf{Input and data-processing:}
\begin{itemize}
    \item 
    \textbf{Users $i$ and events $j$}: $X^C_{i,j},\;i=1,\cdots,N^C,\;j=1,\cdots,M^C_i$, $\sum_{i=1}^{N^C}M^C_i=n^C$; $X^T_{i,j},\;i=1,\cdots,N^T,\;j=1,\cdots,M^T_i$, $\sum_{i=1}^{N^T}M^T_i=n^T$.
    \item
    \textbf{Digits of precision for log-scaling}: $d$, 2 at Snap.
    \item
    \textbf{Unique log-scaled observations and their counts}: $\tilde{X}^C_j,\;j=1,\cdots,K^C$; $\tilde{X}^T_j,\;j=1,\cdots,K^T$. Further let $c^C_i,\;i=1,\cdots,K^C$; $c^T_i,\;i=1,\cdots,K^T$ be their corresponding count in the dataset.
     \item
    \textbf{\# of user buckets}: $s^C=s^T=s$, 100 at Snap.
    \item
    \textbf{\# of bootstrap iterations}: $B^C=B^T=B$, 200 at Snap. 
    \item
    \textbf{Quantile grid locations for QTE evaluation}: $\text{grid}_i\in(0,1),\;i=1,\cdots,N_{\text{grid}}$; e.g., $20\%$ to $99\%$ with step size $1\%$ chosen for Snap.
    \item
    \textbf{Confidence level}: $1-\alpha$, $\alpha=0.05$ at Snap.
\end{itemize}
\textbf{Output: Quantile Treatment Effects--QTE} 
\begin{itemize}
    \item 
    \textbf{$\Delta\%$ and $p$-value evaluated at grid}: $\Delta_i\%,\;p_i,\;i=1,\cdots,N_{\text{grid}}$.
    \item
    \textbf{$\Delta\%$ confidence interval at level $1-\alpha$  evaluated at grid}: $\text{CI}_i:=[\text{CI}_i^L,\text{CI}_i^U],\;i=1,\cdots,N_{\text{grid}}$.
\end{itemize}
for control and treatment identifier $id\in \{C,T\}$, do
\begin{itemize}
    \item evaluate the overall empirical CDF $\mathbb{F}_n^{id}(\tilde{X}_j^{id}),\;j=1,\cdots,K^{id}$. 
    \item split $N^{id}$ users into $s$ buckets of equal size $\floor{N^{id}/s}$; for all users in bucket $i$, put all events in bucket $i$, get corresponding log-scaled event's count $\tilde{X}^{id}_j,\;j=1,\cdots,K^{id}$ as $c^{id}_{i,j}$.
    \item permute the long vector $
\pmb{V}=\left\{\underbrace{1,\cdots,1}_\text{B repetition},\underbrace{2,\cdots,2}_\text{B repetition},\cdots\cdots,\underbrace{s,\cdots,s}_\text{B repetition}\right\}
$ randomly and split into $B$ small vectors $\pmb{v}_i$ of equal length $s$.
    \begin{itemize}
        \item for each bootstrap sample $\pmb{v}_i=(v_{i_1},\cdots,v_{i_s})$, define 
        $\mathbb{F}^{*,id}_{n^{id},i}(x):= \left(\sum_{a=1}^s\sum_{j=1}^{K^{id}}c^{id}_{v_{i_a},j}\mathbb{I}(\tilde{X}_{j}\leq x)\right)/\sum_{a=1}^s\sum_{j=1}^{K^{id}}c^{id}_{v_{i_a},j}$;
        \item
        evaluate $\mathbb{F}^{*,id}_{n^{id},i}(x)$ at $x=\tilde{X}^{id}_1,\tilde{X}^{id}_2,\cdots,\tilde{X}^{id}_{K^{id}}$;
    \end{itemize}
    For each $\tilde{X}^{id}_j,\;j=1,\cdots,K^{id}$, estimate $\text{Var}\left(\mathbb{F}_n^{id}(\tilde{X}^{id}_j)
\right)\approx B^{-1}\sum_{i=1}^B \left(\mathbb{F}^{*,id}_{n^{id},i}(\tilde{X}^{id}_j)-B^{-1}\sum_{k=1}^B\mathbb{F}^{*,id}_{n^{id},k}(\tilde{X}^{id}_k)\right)^2:=\hat{\nu}^B_{n}[\mathbb{F}^{id}_n(\tilde{X}^{id}_j)].$
\item
For each $j=1,\cdots,K^{id}$, let $\hat{\xi}_{\mathbb{F}^{id}_n(\tilde{X}^{id}_j)}^L=\inf\left\{t:\mathbb{F}^{id}_n(t)\geq \mathbb{F}^{id}_n(\tilde{X}^{id}_j)-\hat{\nu}^B_n[\mathbb{F}^{id}_n(\tilde{X}^{id}_j)]\right\}$,
$\hat{\xi}_{\mathbb{F}^{id}_n(\tilde{X}^{id}_j)} ^U=\inf\left\{t:\mathbb{F}^{id}_n(t)\geq \mathbb{F}^{id}_n(\tilde{X}^{id}_j)+\hat{\nu}^B_n[\mathbb{F}^{id}_n(\tilde{X}^{id}_j)]\right\}$, then let 
$\widehat{\text{SE}}(\tilde{X}^{id}_j)=\max\left\{\tilde{X}^{id}_j-\hat{\xi}_{\mathbb{F}^{id}_n(\tilde{X}^{id}_j)}^L,\hat{\xi}_{\mathbb{F}^{id}_n(\tilde{X}^{id}_j)}^U-\tilde{X}^{id}_j\right\}.$
\item
Use linear interpolation on  $\left(\mathbb{F}_n^{id}(\tilde{X}_j^{id}),\widehat{\text{SE}}(\tilde{X}_j^{id})\right),\;j=1,\cdots,K^{id}$, and evaluate on  \text{grid}$_i,\;i=1,\cdots,N_{\text{grid}}\rightarrow \left(\text{grid}_i,\widehat{\text{SE}}\left(\mathbb{F}_n^{id,-1}(\text{grid}_i)\right)\right).$
\end{itemize}
\textbf{QTE}:
\begin{itemize}
    \item \textbf{$\Delta\%$ at grid:}
    $\Delta_i\%=\left(\exp\left(\mathbb{F}_n^{T,-1}(\text{grid}_i)\right)-\exp\left(\mathbb{F}_n^{C,-1}(\text{grid}_i)\right)\right)/\exp\left(\mathbb{F}_n^{C,-1}(\text{grid}_i)\right)\times 100\%;$
    \item \textbf{SE for $\Delta\%$ at grid:}
    $\widehat{\text{SE}}(\Delta_i\%)=\sqrt{\widehat{\text{SE}}\left(\mathbb{F}_n^{C,-1}(\text{grid}_i)\right)^2+\widehat{\text{SE}}\left(\mathbb{F}_n^{T,-1}(\text{grid}_i)\right)^2}\times \frac{\exp\left(\mathbb{F}_n^{T,-1}(\text{grid}_i)\right)}{\exp\left(\mathbb{F}_n^{C,-1}(\text{grid}_i)\right)};$
    \item \textbf{$\Delta\%$ $p$-value at grid:}
    $p_i=2\times\Phi\left(-\left|\frac{\Delta_i\%}{\widehat{\text{SE}}(\Delta_i\%)}\right|\right);$
    \item \textbf{$\Delta\%$ confidence interval at level $1-\alpha$ at grid:}
    $\text{CI}_i=[\Delta_i\%-z_{\alpha/2}\widehat{\text{SE}}(\Delta_i\%),\Delta_i\%+z_{\alpha/2}\widehat{\text{SE}}(\Delta_i\%)].$
\end{itemize}
\end{algorithm*}

\begin{figure}[!h]
    \centering
    \includegraphics[scale=0.38]{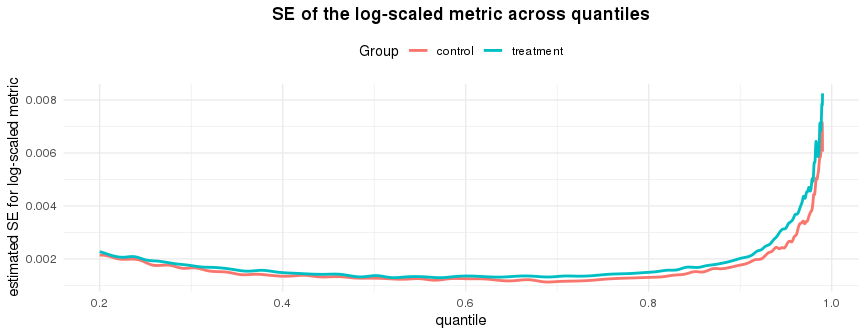}
    \caption{Estimated SE on a grid of quantile locations using Woodruff type estimation mentioned above for control and treatment.}
    \label{fig:est_se_metric}
\end{figure}

Furthermore, using these results joined with the original observations' empirical CDF, we can then linearly interpolate these results on a fixed grid of quantiles (ensuring common quantile locations between control and treatment for QTE evaluation on such grid). At Snap, we chose a range of P20 to P99. The same example referenced in Figure \ref{fig:est_se} has the interpolated standard error, shown in Figure \ref{fig:est_se_metric}.

Once we have such common quantile locations and estimated log-scaled metrics' SE for control and treatment, we have all the ingredients for traditional point-wise $t$-test based statistical inference, including percentage change $p$-values, confidence intervals, etc.

The following section provides a detailed summary of the aforementioned steps constituting the algorithm named CONQ (Continuous QTE for large-scaled experiments).
\subsection{CONQ Algorithm and Implementations}

Here we summarize the end-to-end CONQ algorithm in Algorithm \ref{CONQ}, to adopt earlier notations, superscripts $C$ and $T$ denote data associated with the control and the treatment group respectively, $n$ being the total number of events, $N$ being the total number of users. As can be seen, it only requires several hyper-parameters including $d$ for log-scaling precision, $s$ for number of buckets, $B$ for balanced bootstrap iterations, a pre-determined grid on quantile locations for evaluation, and $1-\alpha$ for confidence level. As for the bag of little bootstrap part, it can be simply construed as re-weighting observations in buckets based on samples on bucket indices. This greatly reduces computational cost in comparison to the method in \cite{Deng18} as the number of unique observations reduce from $n$ to $K$, and $K$ here can be further adjusted by the log-scaling precision $d$.
\begin{figure}[!h]
    \centering
    \includegraphics[scale=0.3]{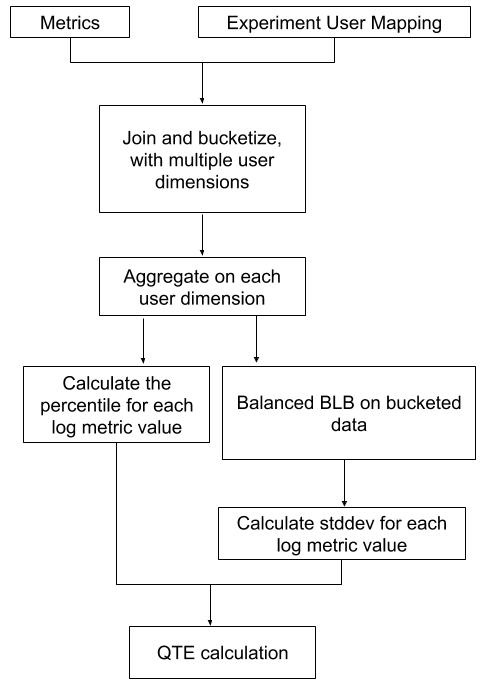}
    \caption{Illustration of CONQ engineering pipeline at Snap, orchestrated through Airflow and Spark.}
    \label{fig:QTE_pipeline}
\end{figure}

One key step that enables us to get a range of continuous QTE evaluation is the evaluation of $\mathbb{F}^{*,id}_{n^{id},i}$ on all $\tilde{X}^{id}_j$ for each bootstrap sample $i$ and $id\in\{C,T\}$. This is straightforward after we do a simple sort on $\tilde{X}^{id}_j$ and a weighted cumulative sum on the re-weighted count $c^{id}_{i,j}$. This avoids 'ntiles' operation and improves efficiency.

We implemented the CONQ algorithm at Snap's A/B experimentation pipeline, the flowchart illustrated in Figure \ref{fig:QTE_pipeline} demonstrates the overall workflow orchestrated by Spark \cite{Zaha16} and Airflow. 

\begin{figure*}
    \centering
    \includegraphics[scale=0.42]{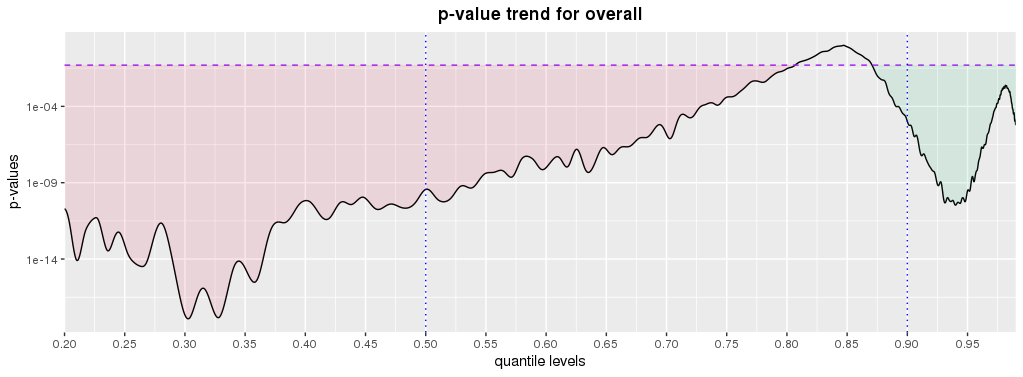}
    \caption{Example in Figure \ref{fig:quantile_example} evaluated by CONQ at P20 to P99, as can be seen, regression (indicated by red) happens mainly at P20 to P80, whereas improvements (indicated by green) mainly happen after P90, indicating presence of heterogeneity in treatment effects. Purple dotted line is plotted at 0.05, y-axis is log-scaled. The threshold for $p$-value here is chosen as 0.05.}
    \label{fig:pval_overall}
\end{figure*}

\begin{figure*}
    \centering
    \includegraphics[scale=0.42]{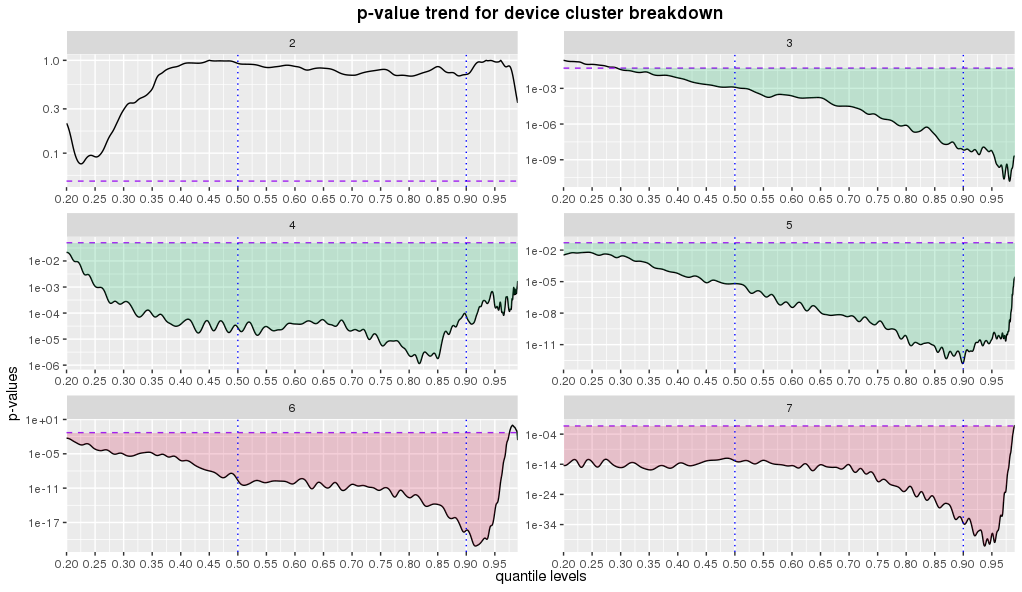}
    \caption{Example in Figure \ref{fig:quantile_example} evaluated by CONQ at P20 to P99 at device cluster breakdown level, as can be seen, regressions mainly occur at high end devices, while improvements mainly occur at low end devices.}
    \label{fig:pval_cluster}
\end{figure*}

Recall our earlier example in Figure \ref{fig:quantile_example} where we see significant but directionally different results at P50 and P90 QTE, here using CONQ, in Figure \ref{fig:pval_overall}, we can see that the regression mainly happens at P20 to P80, i.e. lower quantiles, while improvements mainly happen after P90, thus indicating presence of heterogeneity in treatment effects, especially for devices with varying performance. Further breakdown results by varying device performance clusters (2 to 7, with lower number indicating devices with general poorer performance, while higher number indicating higher performance) in Figure \ref{fig:pval_cluster}, the results corroborate with our speculation that low regressions mainly occur for high end devices, whereas improvements occur for low end devices.

\section{Evaluation}
In this section we evaluate CONQ using real A/B experiments measuring various performance type metrics at Snap. Specifically we aim to address the following questions:
\begin{enumerate}
    \item 
    Comparing with density estimation based methods \cite{Liu19,Deng18}, can CONQ achieve similar results empirically at P50 and P90?
    \item
    Under an known A/A test setting (null case), can CONQ be robust across all quantile locations when it comes to false positives after multiple testing adjustment, thus indicating stability?
\end{enumerate}

\subsection{Evaluation on Snap's A/B Experiments}
For real experiments comparison, we use method in \cite{Deng18} (denoted by \textbf{DELTA}) with kernel density bandwidth choice $h$ chosen by the ad-hoc rule proposed in \cite{Liu19}, and compare $\Delta_\%$ $p$-value results at P50 and P90 with \textbf{CONQ}. A total of 15 performance metrics on 696 treatment and control pairs are evaluated (total of 10440 $p$-values for P50 and P90). The performance metrics included measure a wide array of latencies and crashes for the Snapchat app, which itself should be representative enough for industry A/B experiments.

\begin{figure}
    \centering
    \includegraphics[scale=0.45]{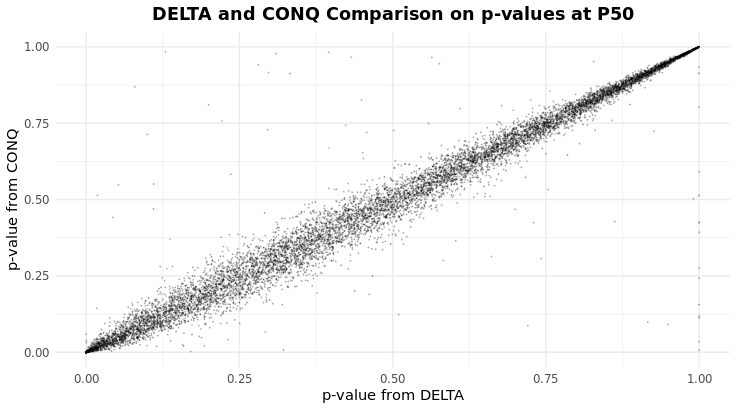}
    \includegraphics[scale=0.45]{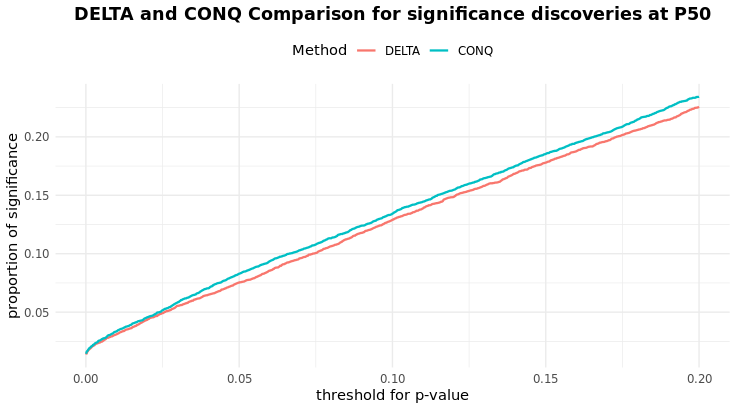}
    \includegraphics[scale=0.45]{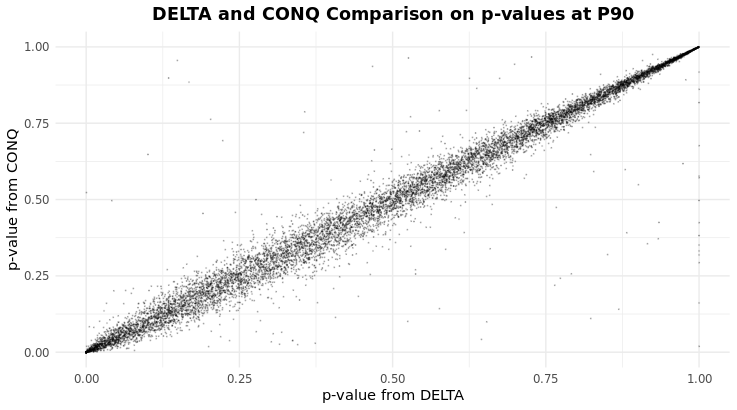}
    \includegraphics[scale=0.45]{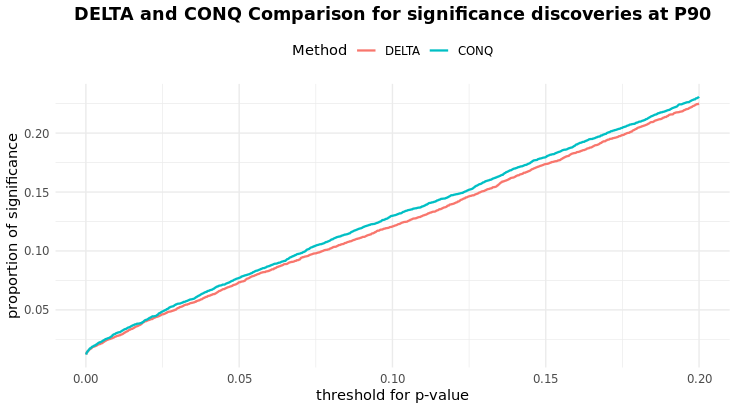}
    \caption{Comparison of DELTA and CONQ on p-values and significance discoveries at various threshold levels.}
    \label{fig:comp}
\end{figure}

We compare the resulting $p$-values from DELTA and CONQ first using a scatter-plot. As can be seen in Figure \ref{fig:comp}, these two methods align quite well and mostly adhere a linear pattern as expected, indicating stable performance of CONQ across various settings and $p$-value regions. Then we compare the proportions of significance discoveries with a static threshold on $p$-values ranging from 0.0001 to 0.2, DELTA and CONQ also perform quite similarly at both P50 and P90 levels, with CONQ resulting in slightly more discoveries across all thresholds.

\subsection{Evaluation on Snap's A/A Tests}
In this section, we evaluate CONQ on a subset of known A/A tests at Snap where no QTE is expected across all quantile locations. Then for any given quantile location, the set of $p$-values associated QTE across different treatment control pairs and various performance metrics should have no significance after multiple testing adjustment using the Benjamini-Hochberg procedure \cite{BH95}. The reason for using BH here instead of testing for a uniform distributed $p$-value, is the positive dependency relationships between many performance metrics, and the robustness of BH under such setting. Here we choose the nominal FDR level as $\alpha=0.05,0.1,0.2$. We evaluate a total of 15 performance metrics across 54 treatment and control pair combinations. The percentile grid for evaluation is chosen as P20 to P95 with $5\%$ increment. BH is applied conditioning on a fixed percentile. Since we know that all scenarios considered are A/A tests, any discoveries by BH are false discoveries. Table \ref{tab:fp} summarizes the percentile and their corresponding count of false discoveries and its percentage (total of 810 $p$-value per quantile). As can be seen, CONQ performs stable across various quantile locations under a realistic FDR nominal thresholds of either $0.05$ or $0.1$. With a larger threshold at $0.2$, CONQ still performs robust with slightly elevated number of false positives in the tail area.
\begin{table}[!h]
\begin{tabular}{|c|c|c|c|}
\hline
\diagbox{\textbf{Percentile}}{\pmb{$\alpha$}} & \textbf{0.05} & \textbf{0.1} & \textbf{0.2} \\ \hline
\textbf{P20}                                          & 0  ($0\%$)           & 0  ($0\%$)          & 0   ($0\%$)         \\ \hline
\textbf{P25}                                          & 0  ($0\%$)           & 0    ($0\%$)        & 0 ($0\%$)           \\ \hline
\textbf{P30}                                          & 0     ($0\%$)        & 0      ($0\%$)      & 0  ($0\%$)          \\ \hline
\textbf{P35}                                          & 0   ($0\%$)          & 0    ($0\%$)        & 0   ($0\%$)         \\ \hline
\textbf{P40}                                          & 0   ($0\%$)          & 0    ($0\%$)        & 0  ($0\%$)          \\ \hline
\textbf{P45}                                          & 0    ($0\%$)         & 0   ($0\%$)         & 0   ($0\%$)         \\ \hline
\textbf{P50}                                          & 0 ($0\%$)            & 0   ($0\%$)         & 0       ($0\%$)     \\ \hline
\textbf{P55}                                          & 0 ($0\%$)            & 0  ($0\%$)          & 0      ($0\%$)      \\ \hline
\textbf{P60}                                          & 0   ($0\%$)          & 0  ($0\%$)          & 0     ($0\%$)       \\ \hline
\textbf{P65}                                          & 0   ($0\%$)          & 0    ($0\%$)        & 0  ($0\%$)          \\ \hline
\textbf{P70}                                          & 0   ($0\%$)          & 0  ($0\%$)          & 0  ($0\%$)          \\ \hline
\textbf{P75}                                          & 0   ($0\%$)          & 0    ($0\%$)        & 0     ($0\%$)       \\ \hline
\textbf{P80}                                          & 0   ($0\%$)          & 0   ($0\%$)         & 1   ($0.12\%$)         \\ \hline
\textbf{P85}                                          & 1   ($0.12\%$)          & 2   ($0.25\%$)         & 3  ($0.37\%$)           \\ \hline
\textbf{P90}                                          & 0    ($0\%$)            & 0 ($0\%$)           & 5    ($0.62\%$)        \\ \hline
\textbf{P95}                                          & 2  ($0.25\%$)           & 2  ($0.25\%$)          & 8     ($0.99\%$)       \\ \hline
\end{tabular}
\caption{\label{tab:fp}: number of false positives discovered and its percentage using various nominal FDR thresholds for BH adjustment out of 810 $p$-values, evaluated at different percentiles. }
\end{table}

\section{Concluding Remarks and Possible Extension}
In this paper, We propose a non-parametric method for point-wise statistical inference on a continuous range of QTE in A/B experiments, which is density estimation free and scalable for the large-scale setting at industry. It achieves stable performance across all ranges and performs similarly to existing delta-method based approaches in \cite{Liu19,Deng18}.

Note the confidence interval provided is conservative to ensure symmetry and $p$-value calculation, for tail areas, the direct application of Woodruff type confidence intervals may have better coverage rate as investigated in \cite{Sitter01}. Furthermore, we can indirectly use the estimated SE on quantile and empirical CDF to get an estimate of the density, which can be used for other tasks, avoiding kernel bandwidth choice. 

we mainly focus on point-wise statistical inference for QTE, namely for any given $p\in(0,1)$, we produce $1-\alpha$ confidence intervals $\text{CI}$ s.t. 
$\mathbb{P}(\xi_p\in\text{CI})\geq 1-\alpha$. We could also be interested in making a confidence interval for all quantiles simultaneously, namely
$\mathbb{P}(\exists p\in(0,1),\;\text{s.t. }\xi_p\in\text{CI})\geq 1-\alpha$, which is in line with the DKWM inequality mentioned above. The method CONQ described in this paper can be further extended to produce $\mathcal{L}_\infty$-norm type bound as well, we can use the bootstrapped variance $\nu$ to get an estimate of the so-called ``effective degrees of freedom", and use it to substitute $n$ in the DKWM inequality or the inequality produced in \cite{Ram19}. This approach is justified by the central limit theory on clustered samples, specifically in \cite{Hansen19}. We leave this for future research.

%%
%% The next two lines define the bibliography style to be used, and
%% the bibliography file.
\newpage
\bibliographystyle{ACM-Reference-Format}
\bibliography{conq}

\appendix

\end{document}